# SIMUS: an open-source simulator for ultrasound imaging.
## Part II: comparison with three popular simulators

Amanda Cigier, François Varray, Damien Garcia

*Background and Objective* – Computational ultrasound imaging has become a well-established methodology in the ultrasound community. In the accompanying paper (part I), we described a new ultrasound simulator (SIMUS) for Matlab, which belongs to the Matlab UltraSound Toolbox (MUST). SIMUS can generate pressure fields and radiofrequency RF signals for simulations in medical ultrasound imaging. It works in a harmonic domain and uses linear equations derived from far-field and paraxial approximations.

*Methods* – In this article (part II), we illustrate how SIMUS compares with three popular ultrasound simulators (Field II, k-Wave, and Verasonics) for a homogeneous medium. We designed different transmit sequences (focused, planar, and diverging wavefronts) and calculated the corresponding 2-D and 3-D (with elevation focusing) RMS pressure fields.

*Results* – SIMUS produced pressure fields similar to those of Field II and k-Wave. The acoustic fields provided by the Verasonics simulator were significantly different from those of SIMUS and k-Wave, although the overall appearance remained consistent.

*Conclusion* – Our simulations tend to demonstrate that SIMUS is reliable and can be used on a par with Field II and k-Wave for realistic ultrasound simulations.

## I. INTRODUCTION

COMPUTATIONAL ultrasound imaging [1] has become popular since 2000 after the availability of the Field II [2] ultrasound simulator (Fig. 1 of Part I). The recent growing interest in computational ultrasound imaging is probably related to the accessibility of raw signals, driven by the emergence of ultrasound open platforms [3]–[5] dedicated to research (e.g. SARUS, UlaOP, Ultrasonix, Verasonics). The interplay between simulations and experimentations has fostered the development and validation of new techniques, both in transmit (e.g. coding, sparse arrays, transmission design) and in post-processing of ultrasound signals (e.g. beamforming, filtering, motion estimation). Access to realistic ultrasound simulations also enables teams not equipped with ultrasound scanners to participate and innovate in the processing of ultrasound signals. The potential offered by ultrasound simulations has motivated some investigators to address configurations not covered by Field II. Other simulators were then created, some of which have been mentioned in Part I. Among the open-source simulators, k-Wave quickly gained popularity and became another reference tool in computational ultrasound imaging. Based on the number of annual citations (Fig. 1 in Part I), k-Wave is now as widely adopted as Field II by the ultrasound imaging community. Although a substantial number of software packages have been developed to broaden the spectrum of simulated ultrasound conditions, Field II and k-Wave are the most widely employed.

In the accompanying paper (part I), we introduced a new ultrasound simulator (SIMUS) that works in the harmonic domain and is written in Matlab language. SIMUS is part of the open-source MUST Matlab UltraSound Toolbox[1] that proposes codes for ultrasound experimentations and simulations. The SIMUS program is based on far-field and paraxial approximations and simulates radiofrequency RF signals for uniform linear or convex probes. SIMUS calls the function PFIELD that returns one-way (transmit) and two-way (transmit + receive) acoustic pressure fields. The theoretical aspect is described in detail in Part I. The RF signals can be analyzed with the open-source functions of the MUST toolbox (beamforming [6], wall filtering, Doppler [7], speckle tracking [8]...) to produce realistic ultrasound images. Some examples are provided in the accompanying paper and on the MUST website.

The objective of this part-II paper was to compare the one-way acoustic pressure fields returned by PFIELD with those simulated by Field II and k-Wave. We also included the Verasonics Research Ultrasound Simulator in the comparison protocol. Although the latter is a recent and therefore little cited [9] software, it could quickly gain popularity as the Verasonics research scanners are widely used by the ultrasound community. Since the theoretical background used in PFIELD/SIMUS is relatively close to that of Field II, except for the fact that SIMUS operates in the frequency domain, we also compared

A.C, F.V. and D. G. are with CREATIS (Centre de Recherche en Acquisition et Traitement de l'Image pour la Santé), CNRS UMR 5220 – INSERM U1294 – Université Lyon 1 – INSA Lyon - Université Jean Monnet Saint-Etienne.

E-mails: francois.varray@creatis.insa-lyon.fr; damien.garcia@creatis.insa-lyon.fr

D.G. is also with INSERM (Institut National de la Santé Et de la Recherche Médicale). E-mails: damien.garcia@inserm.fr; garcia.damien@gmail.com.

[1] https://www.biomecardio.com/MUST

two-way PSFs (point-spread functions) generated by these two software packages. Before describing the comparison protocol and the results, we briefly review the theoretical and numerical specificities of the three simulators Field II, k-Wave, and Verasonics.

## II. Introduction (cont'd): Field II, k-Wave, and Verasonics

Each ultrasound simulator has its purpose and/or numerical specificity. Opting for one rather than the other depends essentially on the targeted goals, but also the ease of use and the realism of the simulations. The four ultrasound simulators that were compared (PFIELD, Field II, k-Wave, Verasonics) can all provide one-way acoustic pressure fields. In this study, we compared RMS (root mean square) acoustic pressures for standard emission sequences. We also compared two-way PSFs (SIMUS vs. Field II) after beamforming. The way PFIELD and SIMUS operate is explained in the part-I paper. Table I summarizes its features and those of the other three simulators.

### A. Field II

J.A. Jensen introduced the Matlab simulator Field II in 1996 [2], after a previous work on the simulation and validation of acoustic pressure fields generated by medical ultrasound transducers [10]. Field II works in the time domain and uses linear acoustics. The acoustic pressure field is calculated by dividing the transducer into small rectangular sub-elements and then summing their individual responses. For a given arbitrary point in the region of interest, Field II calculates the far-field spatial impulse response of a piston-like sub-element [11]. The acoustic pressure field at the same point is then deduced by convolving this local impulse response with the excitation function of the transducer. The received responses are calculated from the acoustic reciprocity principle, by assuming that the point scatterers behave as monopole sources and do not interact with each other (weak scattering).

### B. k-Wave

B.E. Treeby and B.T. Cox introduced the Matlab toolbox k-Wave in 2010 [12], after a previous work on acoustic propagation models in heterogeneous media [13]. The toolbox k-Wave works in the k-space (wavenumber domain). The acoustic field is calculated at the nodes of regular mesh grids using both finite differences (over time) and a pseudo-spectral [14] method (in space). Point monopole sources and sensors are assigned to the grid points that belong to each transducer element. The received responses naturally derive from the numerical acoustic equations included in the software: backscattering occurs whenever an incident wave encounters an acoustic impedance gradient. In contrast to Field II and PFIELD, k-Wave can implicitly handle non-linear propagations and heterogeneous media [15]. Note, however, that some non-linear aspects can be covered with add-ons to Field II [16], [17].

### C. Verasonics

The Verasonics Vantage ultrasound scanners are probably the most widely used research scanners for ultrasound medical imaging. Their flexibility and accessibility to raw ultrasound data make them well suited for 2-D and 3-D ultrasound research. Verasonics recently introduced its ultrasound simulator software. It is likely to become increasingly popular as it is integrated into the Vantage platform, which is why we have chosen to include this simulator in our comparison panel. To date, however, there is no theoretical document describing the Verasonics simulator. According to the Verasonics website [18], the Verasonics simulator simulates backscattered echoes from a set of point targets with a given reflectivity. The contributions of each emitting element are summed by assuming linear acoustic propagation. The transmit waveform, the transmit and receive apodizations, as well as the directivities of the elements are considered to simulate the received RF signals. The website does not specify whether the simulator applies 2-D or 3-D equations. Since the values of the element height and elevation focus did not modify the acoustic pressures, we postulated that the Verasonics simulator is 2-D.

TABLE I
PROPERTIES OF THE FOUR SIMULATORS

|  | PFIELD | Field II | k-Wave | Verasonics |
|---|---|---|---|---|
| numerical method | meshfree | meshfree | mesh | meshfree |
| acoustic equations | linear | linear | nonlinear | linear |
| time domain | harmonic | temporal | both | temporal |
| space domain | $x,y,z$ | $x,y,z$ | $x,y,z + k$ | $x,y$ |
| dimensionality | 2D, 3D | 3D | 2D, 3D | 2D |
| scattering | weak | weak | multiple | weak |
| medium | homogeneous | homogeneous | heterogeneous | homogeneous |
| open-source | yes | no | yes | no |
| free | yes | yes | yes | no |

TABLE II
SIMULATED CONDITIONS (ONE-WAY)

|  | 0°-plane wave | 10°-plane wave | diverging wave | focused wave |
|---|---|---|---|---|
| center frequency | 7.6 MHz | | 2.7 MHz | |
| elements | 128 | | 64 | |
| kerf width | 30 µm | | 50 µm | |
| element width | 0.27 mm | | 0.25 mm | |
| pitch | 0.3 mm | | 0.3 mm | |
| element height | 5 mm (if 3-D) | | 1.4 cm (if 3D) | |
| elevation focus | 1.8 cm (if 3-D) | | 6 cm (if 3D) | |
| 2-way bandwidth | 77 % | | 74 % | |
| TX focus | ∞ | | (0,-1.64) cm | (3,5.2 cm) |
| TX apodization | boxcar | | Tukey | boxcar |

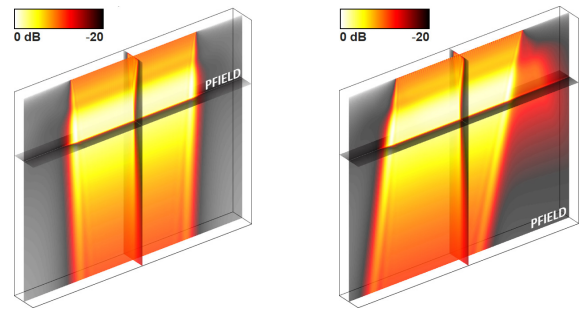

Fig. 1. Simulations of plane waves (tilt angles = 0°, 10°) transmitted by a 128-element linear array with PFIELD (see Table II). The maps represent the RMS (root mean square) acoustic pressure fields. They are 5-cm high.

### III. Method: comparison of the four simulators

#### A. One-way configurations: acoustic pressure fields

We compared the four ultrasound simulators PFIELD, Field II, k-Wave, and Verasonics, by simulating acoustic pressure fields with different emission patterns: plane waves with a tilt angle of 0° or 10°, diverging wave, and focused wave. The properties of the transducers and transmits are summarized in Table II. We tested either 2-D or 3-D conditions or both, depending on the simulator characteristics (Table II). The medium was homogeneous and non-attenuating with a speed of sound of 1540 m/s. The excitation pulse was of one wavelength. The transmit pulse generated by PFIELD was used as pressure input to the other simulators. We simulated the RMS (root mean square) acoustic pressure fields and compared them on a decibel dB scale. The pressure points were located on Cartesian grids with a resolution of one wavelength in all directions. To compare PFIELD with the other simulators, we calculated the absolute differences in the dB scale as $20\log(|P(\text{PFIELD}) - P(\text{other})|)$, after pressure normalization. To obtain a global comparison, we pooled all simulated pressure field values from the same simulator (for 2-D and 3-D, respectively), and calculated the inter-simulator correlation coefficients. To reduce biased trends induced by the lowest values, only acoustic pressures above -20 dB were taken into account in the calculation of the correlation coefficients.

#### B. Two-way configurations: PSF

SIMUS simulates radio-frequency RF signals from the function PFIELD. We compared SIMUS and Field II in two-way configurations by calculating two-way PSFs (point-spread functions). A two-way PSF was defined as the real envelope returned by a single scatterer after beamforming. We tested two ultrasound 3-D conditions: an unsteered plane wave with a linear array (Table II, leftmost column) and a focused wave with a phased array and a transmit focus at (0,6 cm) (Table II, rightmost column). To investigate the two-way PSFs, one scatterer was located at (0,2 cm) and (0,6 cm) in the plane- and focused-wave configurations, respectively. The simulated RF signals were I/Q demodulated and beamformed with a delay-and-sum (DAS) [6]. During beamforming, we used a receive f-number of 1.5 with the linear array, while a full aperture was used with the phased array. We calculated the absolute differences of the PSF in the dB scale as $20\log(|\text{PSF}(\text{SIMUS}) - \text{PSF}(\text{Field II})|)$.

### IV. Results

Fig. 1 to Fig. 7 show the simulated acoustic pressure maps and the absolute differences with respect to PFIELD. We observed a good concordance between PFIELD, Field II, and k-Wave with the 3-D simulations. There was also a good concordance between PFIELD and k-Wave under 2-D conditions. The 2-D acoustic pressure fields simulated by the Verasonics ultrasound simulator were significantly different from those generated by PFIELD and k-Wave. In 3-D, when considering all simulations, the squares of the inter-simulator correlation coefficients were all greater than 0.97 (PFIELD vs. Field II vs. k-Wave, Fig. 8). The correlation between PFIELD and k-Wave was lesser in 2-D ($r^2 = 0.89$), as side lobes were more visible with PFIELD than with k-Wave (Fig. 4, bottom). The 2-D correlation coefficients associated with the Verasonics simulator were the smallest ($r^2 < 0.77$).

The two-way PSFs are displayed in Fig. 9 and Fig. 10. The PSFs returned by SIMUS and Field II were very similar in both plane-wave and focused-wave configurations.

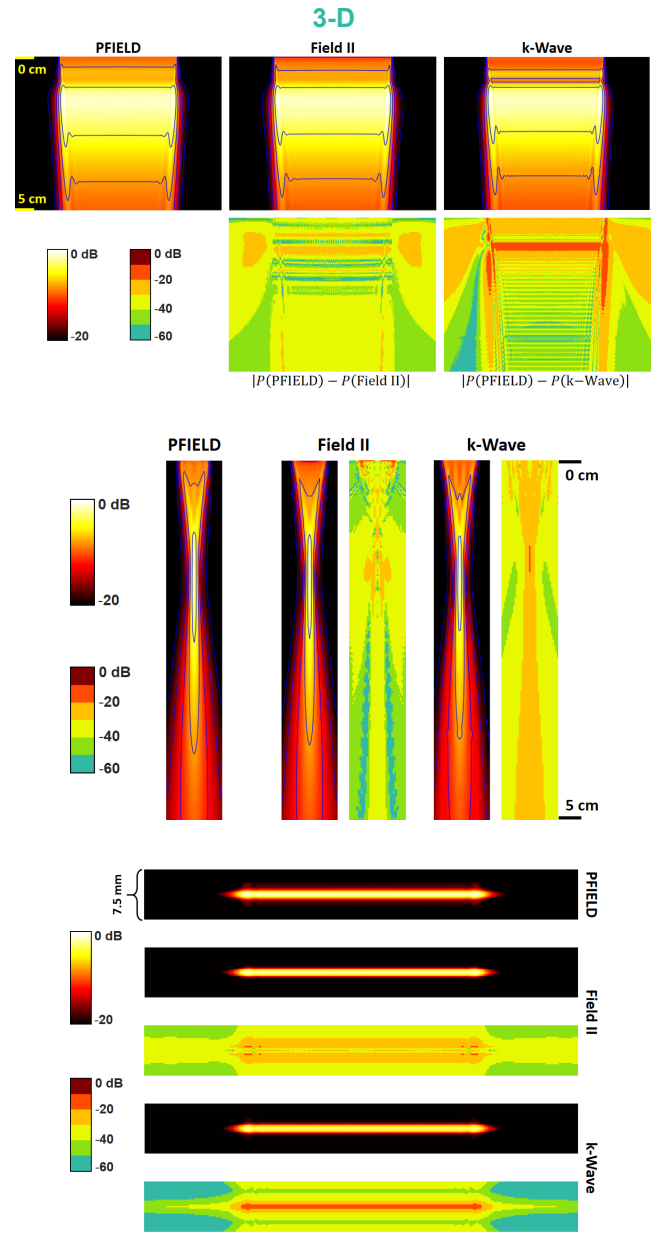

Fig. 2. 3-D simulations: comparison between PFIELD, Field II, and k-Wave for the 0°-tilt plane wave (see Fig. 1, left). The three planes are those displayed in Fig. 1, left. From top to bottom: azimuthal plane, elevation plane, focal plane. The green-to-red maps represent the absolute differences with respect to PFIELD.

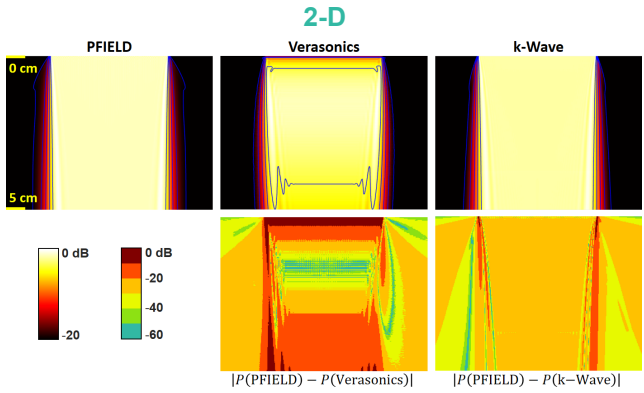

Fig. 3. 2-D simulations: comparison between PFIELD, Verasonics, and k-Wave for the 0°-tilt plane wave. The second row represents the absolute differences with respect to PFIELD.

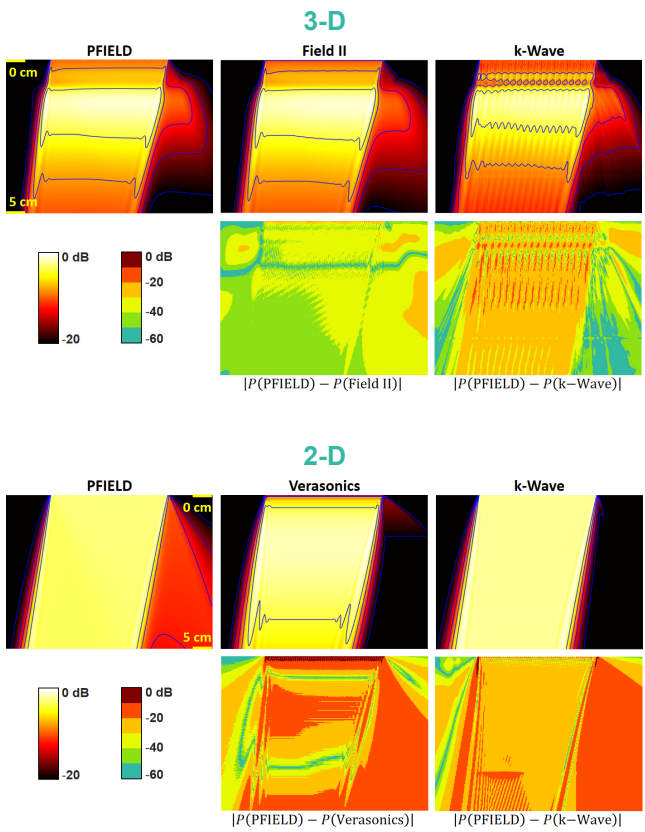

Fig. 4. 3-D (top) and 2-D (bottom) simulations of a 10°-tilt plane wave. A 3-D overview is given in Fig. 1, right. The second rows represent the absolute differences with respect to PFIELD.

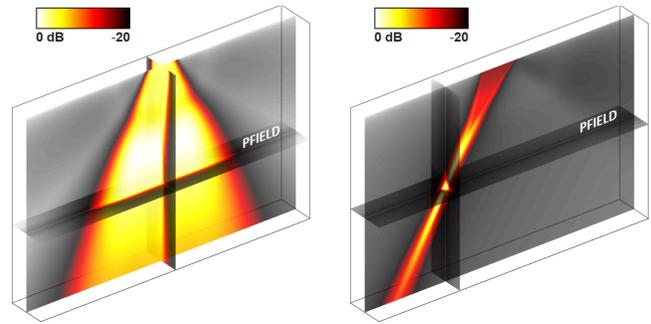

Fig. 5. Simulations of diverging and focused waves transmitted by a 64-element phased array with PFIELD (see also Table II). The maps represent the RMS (root mean square) acoustic pressure fields. They are 10-cm high.

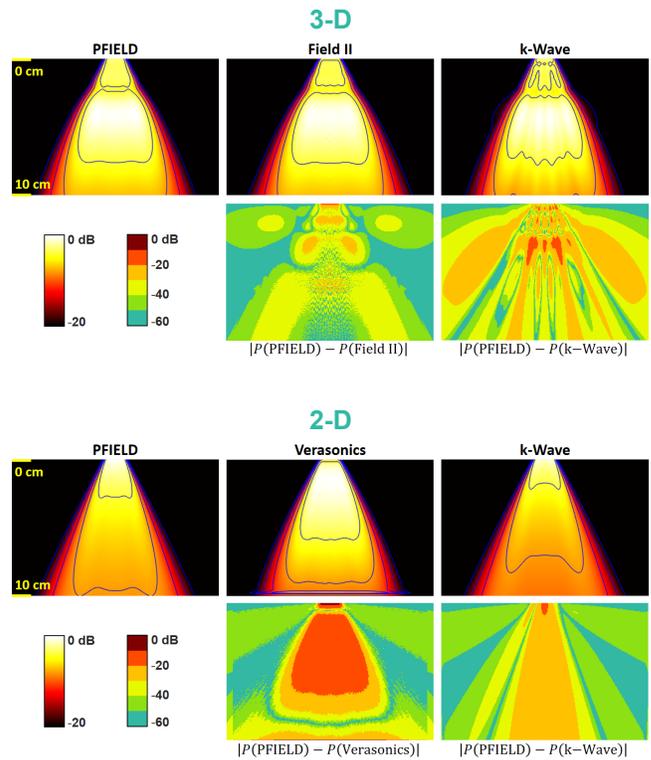

Fig. 6. 3-D (top) and 2-D (bottom) simulations of a diverging wave. A 3-D overview is given in Fig. 5, left. The second rows represent the absolute differences with respect to PFIELD.

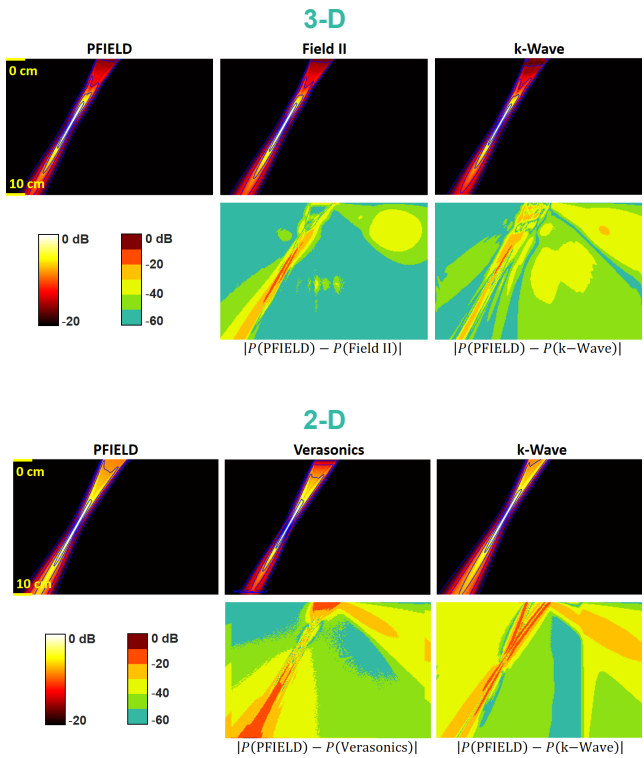

Fig. 7. 3-D (top) and 2-D (bottom) simulations of a focused wave. A 3-D overview is given in Fig. 5, right. The second rows represent the absolute differences with respect to PFIELD.

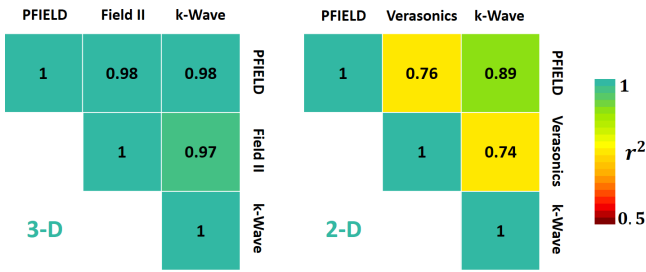

Fig. 8. Overall comparison between the four simulators after pooling all simulated acoustic pressures. The numbers are the correlation coefficients squared ($r^2$).

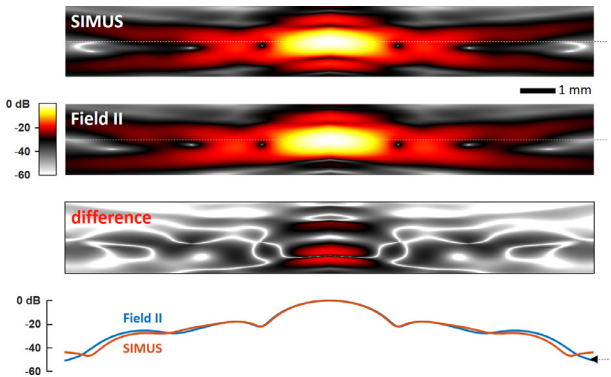

Fig. 9. Comparison of two-way PSFs (SIMUS vs. Field II) in a focused-wave configuration with a simulated phased array.

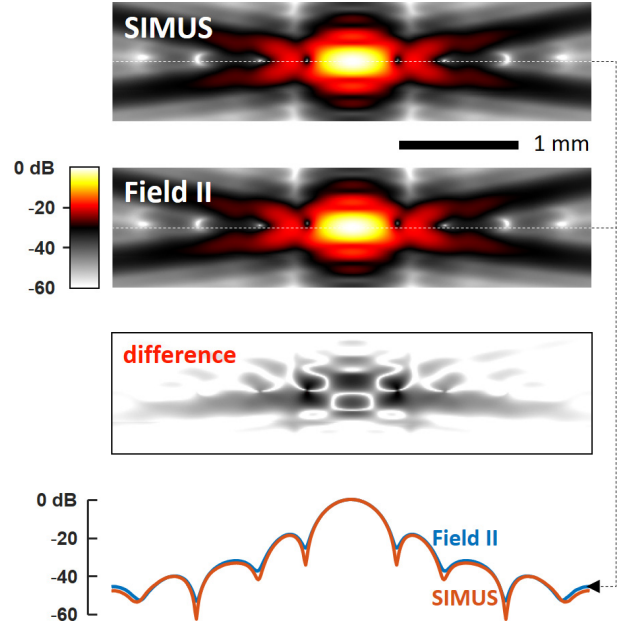

Fig. 10. Comparison of two-way PSFs (SIMUS vs. Field II) in a plane-wave configuration with a simulated linear array.

## V. Discussion

The new ultrasound simulator SIMUS generated acoustic pressure fields similar to those simulated by the most popular software packages Field II and k-Wave, especially in 3-D ($r^2 > 0.96$, Fig. 8). Under 2-D conditions, although the radiation patterns appeared to be identical, we observed some differences with the Verasonics simulator ($r^2 < 0.77$, Fig. 8). Our simulations tend to demonstrate that SIMUS is reliable and can be used on a par with Field II and k-Wave for realistic ultrasound simulations. Regarding the differences found with the Verasonics commercial simulator, it is difficult to give an opinion because, unlike Field II and k-Wave, the theoretical and numerical methods are not publicly available. The best match was with Field II. This is unsurprising because we founded SIMUS based on the same assumptions as Field II, namely: linearity, homogeneity, and far-field approximations with the use of sub-elements. Like Field II, SIMUS uses weak and monopole scattering. It follows that the two-way PSFs obtained with SIMUS and Field II were very similar. The main numerical difference between PFIELD/SIMUS and Field II is the computational domain: temporal with Field II, frequency with SIMUS. Another difference appears with regard to elevation focusing. In SIMUS, to simulate focusing in the elevation direction, we adjusted the velocity delays of the pistons along their height (see Fig. 4 in the accompanying part-I paper), as suggested in [19]. In Field II, the elements are split into sub-elements in the height direction. A transmission delay is associated with each of these sub-elements to provide focusing in elevation. A similar strategy is used in k-Wave. Given the theoretical similarities between PFIELD and Field II, it is legitimate to wonder why we propose yet another simulator.

*A. When using SIMUS and PFIELD?*

SIMUS and PFIELD are part of the Matlab Ultrasound Toolbox MUST toolbox. This toolbox, as illustrated by the numerous examples available on the website[2], allows students and researchers to quickly learn experimental ultrasound imaging: acquisitions, demodulation, beamforming, wall filtering, color Doppler, vector Doppler, speckle tracking… SIMUS and PFIELD complete the series by offering to perform realistic ultrasound simulations for pedagogical and research purposes. In contrast to FIELD II, the Matlab codes are open and highly documented. The default functions allow many tasks to be performed, such as those illustrated by the examples in the Part-I paper. An advanced user can modify the codes or the programming language according to her needs. PFIELD and SIMUS have been already used in several published works. Far-field patterns were generated with PFIELD to design diverging-wave transmit sequences for high-frame-rate motion-compensated echocardiography in human hearts [20]. Vector flow imaging with vector Doppler was simulated in a carotid bifurcation by coupling SIMUS with an SPH (smoothed particle hydrodynamics) flow model [21]. In our team, SIMUS has also recently been applied in the generation of moving ultrasound phantoms to train convolutional neural networks for motion estimation in ultrasound imaging [22]. We also simulated clinical-type Doppler-echo cineloops with the future objective to train deep learning algorithms for intracardiac flow imaging [23]. In the same vein, Milecki *et al.* trained a convolutional neural network by using SIMUS simulations in a synthetic murine brain microvascular network for deep-learning-based ultrasound localization microscopy [24].

*B. What SIMUS cannot do, what it can*

Alike Field II and k-Wave, SIMUS can be used to produce realistic radiofrequency RF signals for ultrasound imaging. Only linear acoustics into homogeneous media, however, is considered in the current version of SIMUS. The homogeneity hypothesis means that both density and speed of sound are constant. Aberrations due to acoustic impedance gradients thus cannot be simulated. The linearity assumption simplifies the problem since it allows the summation of the time-harmonic solutions, but this prevents one from addressing non-linear properties as in harmonic imaging. SIMUS also considers only weak (single) scattering to avoid computational overload of multiple scattering. The multiple reverberation artifacts, which are sometimes observed in clinical ultrasound, therefore cannot be simulated. In contrast, the simulator k-Wave can account for nonlinearity, acoustic heterogeneities, and multiple scattering. However, because k-Wave uses a mesh-based method, the locations of the scatterers are linked to the geometry of the grid. This represents a major difference with SIMUS (or Field II), which is a particle-based software. In SIMUS (or Field II), the speckles of the B-mode images result from the random positions of the scatterers (with constant acoustic impedance), while they are due to local changes in impedance in k-Wave (with unchanged scatterers' positions). In addition to numerical simplifications, the use of a particle-based model readily allows the motion of individual particles. This makes it an ideal simulation method for tissue displacements and the investigation of ultrasound techniques for motion analysis (e.g. speckle tracking [22], Doppler [25], vector flow imaging [21]).

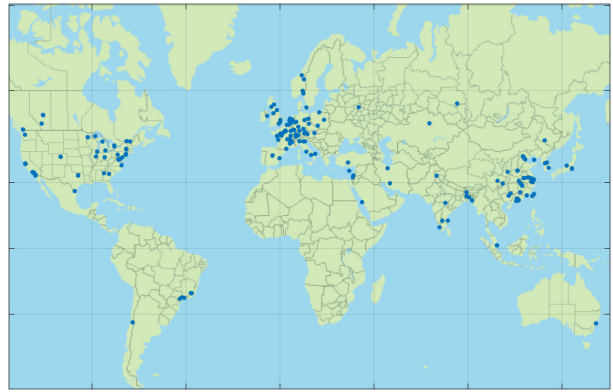

Fig. 11. Cities where SIMUS has been used before October 2021.

*C. Potential future perspectives*

Since the MUST toolbox is available online, PFIELD and SIMUS have been used or tested by researchers and students on several continents (Fig. 11). Depending on the success of SIMUS, we will incorporate the option to simulate matrix arrays for volume ultrasound imaging. It is understood that this will significantly increase the number of element-scatterer interactions. Some numerical strategies will have to be implemented to reduce the computation time. As we are interested in the simulation of cardiac images (echocardiography), we may also plan to integrate an add-on for the calculation of the second harmonic using a quasi-linear approximation of the nonlinear wave propagation [26], [27]. Deep learning is gaining momentum in ultrasound imaging. We anticipate that ultrasound simulators still have a bright future ahead of them as they can be used to train neural networks with limited effort.

VI. CONCLUSION

The open-source Matlab simulator SIMUS is efficient and easy to use. Like Field II and k-Wave, SIMUS could contribute to the growing popularization of computational ultrasound imaging.

ACKNOWLEDGMENT

This work was supported in part by the LABEX CeLyA (ANR-10-LABX-0060) of Université de Lyon, within the program "Investissements d'Avenir" (ANR-16-IDEX-0005) operated by the French National Research Agency.

---

[2] https://www.biomecardio.com/MUST